# Giant Phononic Anisotropy and Unusual Anharmonicity of Phosphorene: Interlayer Coupling and Strain Engineering

Yongqing Cai[†], Qingqing Ke[‡], Gang Zhang[*,†], Yuan Ping Feng[§],

Vivek B Shenoy[¶], and Yong-Wei Zhang[*,†]

[†]Institute of High Performance Computing, A*STAR, Singapore 138632
[‡]Department of Materials Science and Engineering, National University of Singapore, Singapore 117574
[§]Department of Physics, National University of Singapore, Singapore 117542
[¶]School of Engineering & Applied Science, University of Pennsylvania, Philadelphia, PA 19104-6391

**ABSTRACT:** Phosphorene, an emerging elemental two-dimensional (2D) direct band gap semiconductor with fascinating structural and electronic properties distinctively different from other 2D materials such as graphene and MoS2, is promising for novel nanoelectronic and optoelectronic applications. Phonons, as one of the most important collective excitations, are at the heart for the device performance, as their interactions with electrons and photons govern the carrier mobility and light-emitting efficiency of the material. Here, through a detailed first-principles study, it is demonstrated that monolayer phosphorene exhibits a giant phononic anisotropy, and remarkably, this anisotropy is squarely opposite to its electronic counterpart and can be tuned effectively by strain engineering. By sampling the whole Brillouin zone for the mono-layer phosphorene, several "hidden" directions are found, along which small-momentum phonons are "frozen" with strain and possess the smallest degree of anharmonicity. Unexpectedly, these "hidden" directions are intrinsically different from those usually studied armchair and zigzag directions. Light is also shed on the anisotropy of interlayer coupling of few-layer phosphorene by examining the rigid-layer vibrations. These highly anisotropic and strain-tunable characteristics of phosphorene offer new possibilities for its applications in thermal management, thermoelectronics, nanoelectronics and phononics.

## 1. Introduction

Research on two-dimensional (2D) materials, such as graphene, hexagonal BN (h-BN) and MoS$_2$, has been greatly increased owing to their distinct electronic and structural properties, which enable the development of novel nanoelectronic devices.[1-3] Recently, another atomically thin elemental 2D material, phosphorene, formed by mechanically cleaving the layered black phosphorus (BP), has attracted increasing attention[4-12] due to its sizeable band gap (~1.5 eV for mono-layer phosphorene)[13] and high mobility (~1000 cm$^2$v$^{-1}$s$^{-1}$ at room temperature).[14] Similar to graphene, phosphorene has a honeycomb structure, which is, however, strongly puckered with ridge and accordion structures along the zigzag and armchair directions, respectively (refer to **Figure 1a**). The recently surged research interest in phosphorene is partially attributed to its highly puckered structure which leads to intriguing electronic and optical properties.[14-20] Notably, a large hole mobility and an orientation-dependent charge transport behavior were observed for a 10 nm thick BP film.[21] Angle-resolved Hall mobility measurement showed about 1.8 times higher mobility along the accordion direction than along the ridge direction.[22] Applying external stimulus such as strain[23-27] and electric fields[28-29] was demonstrated to be an effective approach to modulate its electronic structure. In principle, the intrinsic scattering of carriers in layered nanomaterials is largely attributed to phonons,[30,31] including acoustic phonons and low-frequency rigid modes, with the latter being exclusively found in layered materials.[32-34] Moreover, phonon-phonon scattering determines the thermal transport properties of 2D materials, which often show a significant size-dependence.[35-38] Very recently, the thermoelectric property and thermal conductance of strainless monolayer phosphorene was studied and a high value of the figure of merit was predicted.[39] The lattice vibrational modes and corresponding Raman spectra of strained monolayer phosphorene were studied and compared favorably with experimental results.[40] However, some phononic properties of phosphorenes are still largely unknown, such as properties of low frequency acoustic phonons, Grüneisen parameter, and phonon behavior of multilayer phosphorene, despite their important roles in thermal management and carrier scattering.

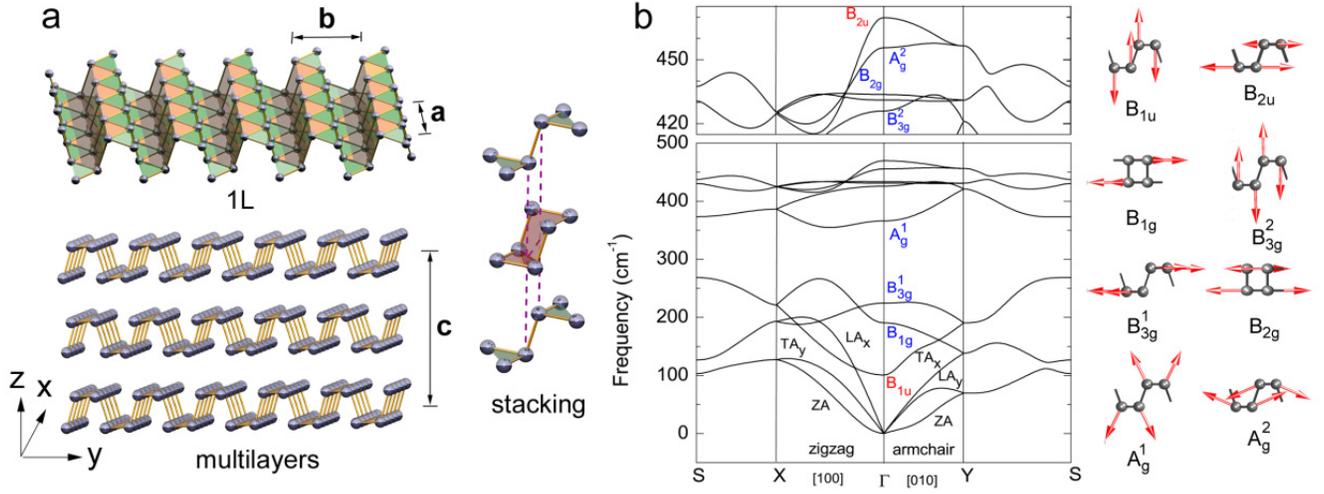

**Figure 1**. Atomic model and phonon dispersions of phosphorene. **a,** Ball-stick models of mono-layer and few-layer phosphorenes. The puckered hexagon formed by six-phosphorus atoms, and the AC-stacked few-layer phosphorene are shown in the inset. **b,** Phonon dispersion and assignment of IR and Raman modes of phosphorene (bottom left). A suffix to the LA and TA acoustic branches denotes the polarization direction of the modes. A close-up view of the high energy optical modes (top left). The vibrating patterns of the optical modes are shown in the right side.

In this work, by performing density functional theory (DFT) and density functional perturbation theory (DFPT) calculations, we investigate the atomic vibrations in mono- and few-layer phosphorene and the inherent structural and phononic responses of mono- and few-layer phosphorene by modulating in-plane mechanical deformation. A strong phononic anisotropy, which is squarely opposite to its electronic counterpart, is found in phosphorene. Importantly, through fully sampling the 2D Brillouin zone, we show that long wavelength acoustic phonons propagating along several special directions possess a much smaller anharmonicity than along the armchair and zigzag directions. By calculating the low-frequency in-plane shearing and out-of-plane breathing modes in few-layer phosphorene, we find a strong anisotropy in the interlayer coupling, which suggests a high orientation-dependent thermal transport in few-layer phosphorene.

## 2. Methodology

We have calculated the interatomic force constants (IFC) and phonon dispersion by DFT and DFPT calculations using the Quantum-Espresso code.[41] Norm-conserving local density approximation (LDA) functional of Perdew and Wang with energy cutoff of 50 Ry is adopted. The first Brillouin zone is sampled with a $14 \times 10 \times 4$ Monkhorst-Pack (MP) mesh for bulk BP and a $14 \times 10 \times 1$ MP mesh for few-layer phosphorene which is modelled using a supercell slab with periodic replicas along the normal direction separated by a 16 Å vacuum region. The structures and the in-plane lattice constant of few-layer phosphorene and bulk BP are relaxed until the forces exerted on the atoms are less than 0.01 eV/Å and the stress exerted on the cell is less than 0.1 kbar. Comparison is also made with calculations using optimized Becke88 (optB88)[42] by Vienna ab initio simulation package (VASP)[43] with a cutoff energy of 500 eV. For the structure under uniaxial strains, the lattice normal to the strain direction is also relaxed. Based on the fully relaxed structures, the dynamic matrix is calculated at **q** points defined in an $8 \times 6 \times 1$ MP mesh, and the IFC is obtained through inverse Fourier transform to real space.

## 3. Results

The initial few-layer phosphorene structures, represented by $n$L, where $n$ is the number of layers, are adopted from the bulk BP determined by experimental measurement.[44] The bulk BP has an orthorhombic symmetry consisting of double puckered layers, which are coupled by van der Waals (vdW) interactions. As shown in Figure 1a, the ridge structures are formed in each phosphorene sheet by the shared edges of neighboring hexagons. The formation of the strongly puckered hexagonal rings is attributed to the favored $s^2p^3$ hybridized states of the phosphorus atoms, which are different from the $sp^2$ states of carbon atoms in graphene although both structures have the same coordination number of three.

The phonon dispersion of mono-layer phosphorene is shown in Figure 1b. With four atoms in the unit cell, phosphorene has three acoustic branches: the in-plane transverse acoustic (TA) mode, the longitudinal acoustic (LA) mode and the out-of-plane acoustic (ZA) mode, and nine optical phonon branches, which are labeled according to the irreducible representation of modes at the $\Gamma$ point, same as bulk BP:[45-48] the two IR-active modes ($B_{1u}$ and $B_{2u}$), six Raman-active modes ($A_g^1$, $A_g^2$, $B_{1g}$, $B_{2g}$, $B_{3g}^1$, and $B_{3g}^2$) and one silent mode ($A_u$). To validate the accuracy of our methods, the calculated frequencies of these optical modes by using both the LDA and the optB88[42] functionals are listed in **Table 1**, which are in good agreement with previous studies.[21,40] The LDA calculations predict very accurate frequencies for the Raman-active $A_g^1$ (366.1 cm$^{-1}$), $B_{2g}$ (433.7 cm$^{-1}$), and $A_g^2$ (455.6 cm$^{-1}$) modes, compared respectively favorably with the experimental values of 360.6, 438.9 and 469.0 cm$^{-1}$.[21,49] Clearly, the optB88 method underestimates the values of these frequencies. While the vdW corrections improve the lattice parameters over LDA and GGA in comparison with experimental results,[50] the improvement in predicting the frequencies by the vdW corrections is less pronounced. The underlying reason for this is because the calculation of the vibrational properties requires a secondary derivative of the total energy, thus the contribution of vdW interaction is negligible. The present results are consistent with previous studies on graphite and $MoS_2$ which showed that LDA yields not only accurate phonon dispersions but also the frequencies of those flexural modes for layered structures.[32,34] Therefore, in the following sections in deriving the lattice dynamics properties, only the LDA approach is adopted.

As shown in Figure 1b, the LA and TA modes around the zone center along the accordion (armchair) direction show a lower group velocity than those along the zigzag direction. In addition, the phonon bands of the optical modes are overall more flat along the accordion direction than along the ridge direction. These relatively localized branches suggest that the vibrational atoms are somehow weakly coupled along the armchair direction. By contrast, the electronic structure shows a completely opposite character with clearly more dispersive bands along the armchair direction than the zigzag direction.[16] Therefore, from the vibrational point of view, we can describe phosphorene as a 2D puckered honeycomb structure consisting of weakly coupled quasi-one dimensional (1D) zigzag chains formed in the ridge direction. Hence, the strong anisotropy of phononic properties in phosphorene revealed here is intrinsically different from its electronic properties, enabling the asymmetrical transport of electrons/holes and phonons[51,52] and offering new possibility for novel low-dimensional thermoelectrics devices. In addition, the size-, shape- and temperature-dependent characteristics of thermal conductivity in nanomaterials offer plenty of opportunities for phonon engineering of phosphorene by exciting different types of phonons.

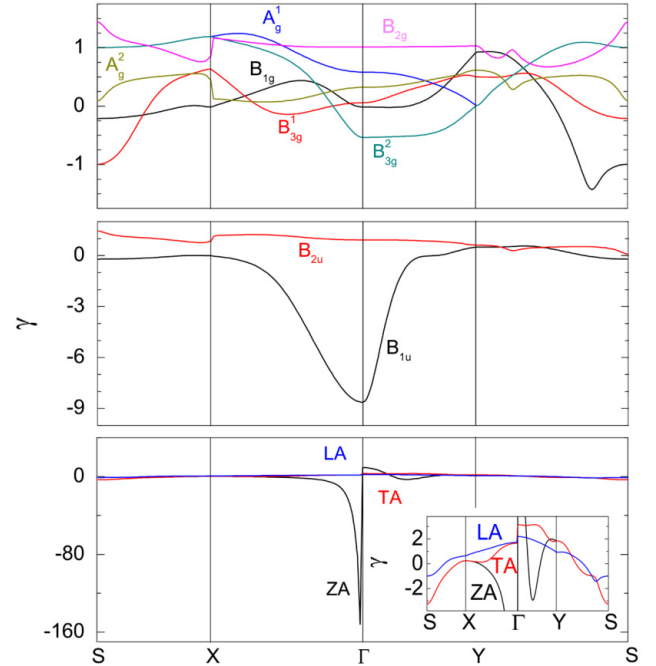

**Figure 2**. Grüneisen parameter and evolution of frequencies under strain. Momentum- and polarity-dependent Grüneisen parameter (γ) of phonon branches with Raman (upper) and IR (middle) active modes at Γ-point, and the acoustic modes (bottom) along the S-X-Γ-Y-S high symmetric path.

We next estimate the anharmonicity of phonons, which reflects the strength of the phonon-phonon scattering for each branch. In general, Grüneisen parameter (γ), which is momentum- (*q*) and polarity- (*s*) dependent, is a good indicator of the anharmonicity of the interatomic potential, and can be calculated based on the biaxial strain using the following formula:

$$\gamma_{\mathbf{q}s} = -\frac{a}{2\omega_s(\mathbf{q})}\frac{d\omega_s(\mathbf{q})}{da} \qquad (1)$$

where *a* is the relaxed equilibrium lattice constant. **Figure 2** shows $\gamma_{\mathbf{q}s}$ within the Brillouin zone for all the Raman, IR, and acoustic modes. The exact values of γ for all the optical modes at Γ are compiled (refer to Table 1). For the Raman modes, the values of γ are found to vary between -1 to 1, and a negative γ is found for the $B_{1g}$ and $B_{3g}^2$ modes. Overall, $\gamma_{\mathbf{q}s}$ along the Γ-Y branch is less dispersionless compared to the Γ-X direction, similar to the phonon dispersion case (Figure 1b). For the IR modes, the value of γ for the low frequency $B_{1u}$ mode is -8.63, whereas the B$_{2u}$ mode has a positive value of 0.92. Notably different from the isotropic 2D sheet like graphene[53] and MoS$_2$,[54] phosphorene exhibits a strongly anisotropic shape of $\gamma_{qs}$ around Γ between the X-Γ and Γ-Y branches. For ZA branch, the value of $\gamma$ near the Γ-point from the X-Γ part is around -160 while it becomes around 10 when approaching from the Γ-Y side, implying that there is a large difference in thermal expansion along the zigzag and armchair directions with decreasing temperature. This strong anisotropic behavior also suggests an orientation-dependent anharmonicity and a different degree of phonon-related scattering for electrons and phonons between the two directions.

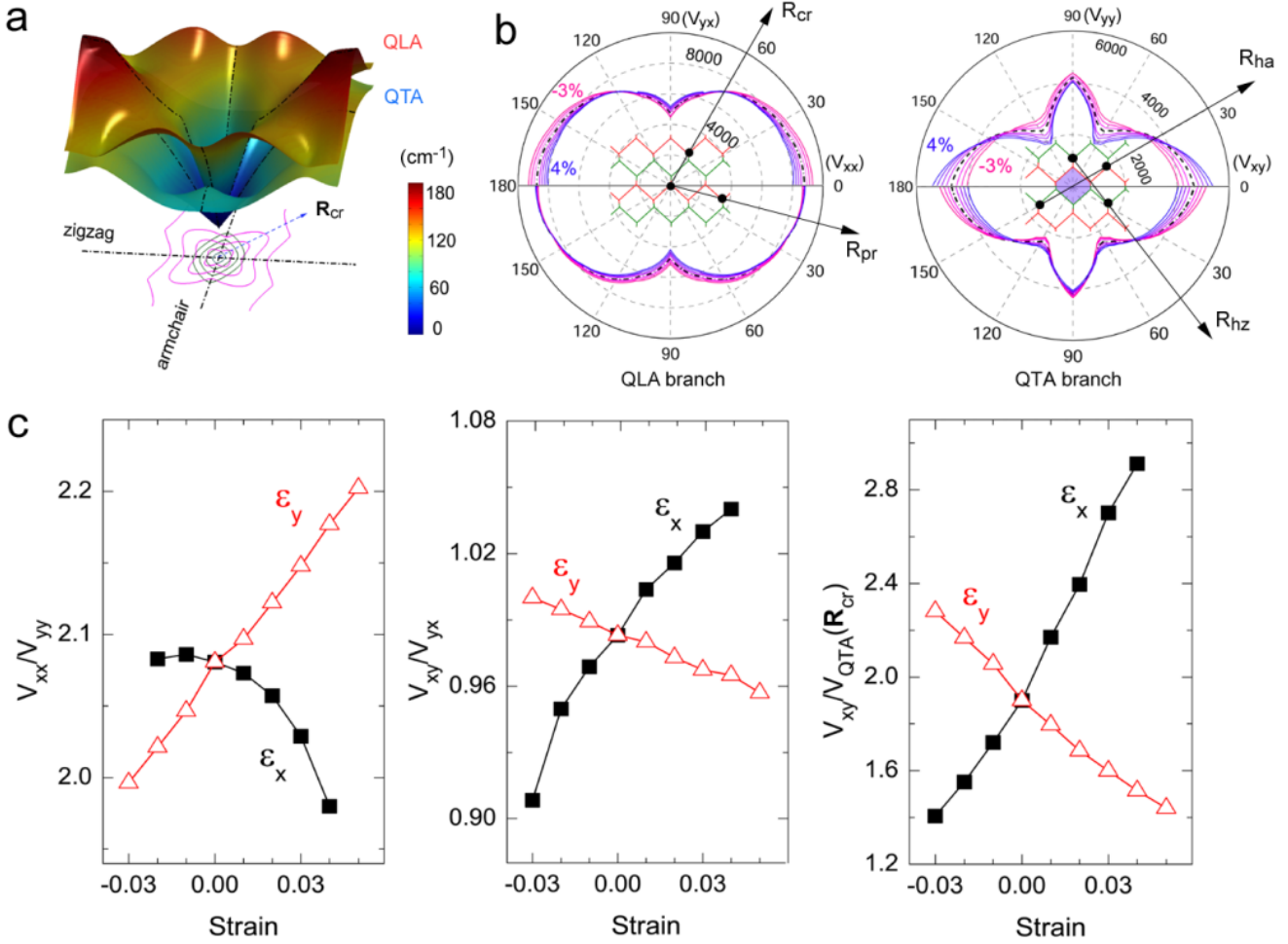

**Figure 3**. Anisotropy of the acoustic phonons. **a,** Phonon dispersion surfaces for the in-plane acoustic branches. The green (pink) lines are the contour plots of QLA (QTA) branches with an increment of the frequency of 20 cm$^{-1}$. The black dash-dotted lines in zigzag and armchair directions are guides to the eye. The dashed blue arrow indicates $R_{cr}$ with the softest in-plane shear deformation (see below). **b,** Orientation-dependence of the group velocity (in units of m/s) around Γ as a function of uniaxial $\varepsilon_x$ (upper part) and $\varepsilon_y$ (lower part) strains from -3% to 4% for the QLA (left) and QTA branches (right). The polar directions at 0 and 180° correspond to the zigzag direction and 90° the armchair direction. Dashed lines correspond to the strain-free phosphorene. Note the anomalous character of $V_{xy}$ for the TA branch under the $\varepsilon_x$ strain. **c,** The variation of the degree of the anisotropy of the velocity with strain defined as the ratio of the velocity along the zigzag direction and that along the armchair direction for the LA (left) and TA (middle) branches. The highest anisotropy (right) for TA branch is obtained along the zigzag and $R_{cr}$ directions.

Since the acoustic modes play a critical role in umklapp scattering of phonons for limiting thermal conductivity, we next investigate the anisotropy of the acoustic phonon dispersion and its strain modulation of phosphorene. In principle, under the theoretical regime of relaxation time approximation, the contribution from the out-of-plane ZA mode to the thermal conductivity in layered materials is much less than those from LA and TA modes due to the small group velocity.[55] Therefore, here, we focus on the in-plane acoustic phonon modes. **Figure 3a** shows the dispersion surfaces of the two in-plane branches within the first Brillouin zone. We define the high-frequency and low-frequency branches as quasi-LA (QLA) and quasi-TA (QTA) instead of LA and TA modes, respectively, considering the fact that most of the in-plane acoustic modes inside the Brillouin zone have a mixing character of transverse and longitudinal vibrations (pure LA and TA modes only exist respectively in the zigzag and armchair directions). Moreover, along the Γ-Y (armchair) direction, as shown in Figure 1b, the $TA_x$ branch has a higher energy than the $LA_y$ mode, being opposite to the common notion that TA modes tend to be softer. We also show the contour plots with a frequency increment of 20 $cm^{-1}$ for the QLA and QTA branches around the zone center in Figure 3a, whereby a strong anisotropy of the dispersion is clearly visible for both branches.

In Figure 3b, the orientation-dependent group velocity ($V$) for the small-momentum acoustic phonons, defined as $d\omega/dq$, is plotted by calculating the slope of the dispersion surfaces around Γ. Hereafter, we use $V_{xy}$ to denote the velocity of phonons propagating along **x** direction and polarized along **y** direction. Under zero strain, the value of $V_{xx}$ for QLA branch ($LA_x$ mode along the zigzag direction) is 8762 m/s, about two times of $V_{yy}$ (4211 m/s) for $LA_y$ mode along the armchair direction. In contrast, the $V_{xy}$ and $V_{yx}$ for the $TA_x$ and $TA_y$ modes, which are 4656 and 4737 m/s, respectively, are almost the same, as required by the stress-free boundary condition ($V_{xy} = V_{yx}$). Along the **y** direction, the $V_{yy}$ of the longitudinal $LA_y$ mode, is comparable to the $V_{yx}$ of the $TA_y$ mode, consistent with the nearly accidentally degenerate modes along the armchair direction around Γ as shown in Figure 1b.

Consistent with the dispersion curve as shown in Figure 1b, within the Brillouin zone, we obtain overall more dispersive acoustic curves and accordingly a faster transport of phonons along the Γ-X (zigzag) direction than along the Γ-Y (armchair) direction. This characteristic is more apparent for the optical modes, especially for the high-frequency modes where bands along the Γ-Y line are relatively more flat than those along the Γ-X line (see Figure 1b). This is in sharp contrast to the electron case, where the electronic bands along the Γ-Y direction are more dispersive.

When a strain is applied, the angular dependence of the acoustic velocity $V$ changes accordingly (Figure 3b). Uniaxial strain ranging from -5% to 5% along the ridge (**x**) and the accordion (**y**) directions, denoted by $\varepsilon_x$ and $\varepsilon_y$ respectively, are considered. For the QLA modes, clearly, the strain $\varepsilon_x$ ($\varepsilon_y$) presents a maximum change of $V$ at 0 (90°), corresponding to the pure longitudinal $LA_x$ (transversal $TA_x$) modes. Both $V_{xx}$ and $V_{yy}$ increase with increasing compressive strains. For the QTA modes, a more significant variation of $V$ is found along the zigzag direction than along the armchair direction under both $\varepsilon_x$ and $\varepsilon_y$ strains. While $V_{xy}$ decreases under tensile strain $\varepsilon_y$, it unexpectedly increases under tensile strain $\varepsilon_x$. Clearly, this anomaly originates from the electron-lattice interaction. As shown in Figure 3c, the uniaxial strains have opposite effect on the anisotropy for the LA and TA modes, which is quantitatively analyzed by calculating the ratio of $V$ along the armchair and zigzag directions. For the LA modes, the tensile strain $\varepsilon_x$ decreases the anisotropy ($V_{xx}/V_{yy}$) while the tensile strain $\varepsilon_y$ increases the anisotropy, whereas for the TA modes, an opposite trend of the anisotropy ($V_{xy}/V_{yx}$) occurs for the strains $\varepsilon_x$ and $\varepsilon_y$.

It is interesting to find that there exist several special directions, as indicated by the arrows in Figure 3b, along which the phonons with a small momentum ($q_s$) are "frozen" with their velocities (equivalently the slope of the dispersion) being kept almost unchanged with strain. To align these directions in the crystallographic coordination, we also plot in Figure 3b the atomic structure of phosphorene in the center of the polar plots. The two atomic ridges in the upper and bottom surfaces are plotted in red and green, respectively. For both QLA and QTA branches, there are two nonequivalent directions of the "frozen" modes, corresponding to $\varepsilon_x$ and $\varepsilon_y$, respectively. For the QLA branch, when the strain $\varepsilon_y$ is applied, one direction is denoted as $R_{pr}$ which is slightly tilted about 15° from the ridge direction. The other direction, which is associated with the strain $\varepsilon_x$ and represented by $R_{cr}$, is along the bonding direction between the two nearest atoms cross the neighboring bottom or upper ridges, and forms around 30° angle with the accordion direction. For the QTA branch, the two directions, labeled as $R_{ha}$ and $R_{hz}$, form angles about 30° and 45° with the ridge, respectively. In the perfect honeycomb case, the $R_{ha}$($R_{hz}$) and armchair (zigzag) direction along the ridge (accordion) direction can be exchanged with each other by a $C_3$ rotation. Rooted in the symmetry reduction from the perfect hexagonal to the orthorhombic structure, the atomic profiles along the two directions in phosphorene become nonequivalent. Here we show that these "hidden" $R_{ha}$ ($R_{hz}$) directions, which were unfortunately ignored in previous studies, should have dramatically different properties, such as smaller phonon deformation potential and significantly less anharmonicity, from the widely investigated armchair and zigzag directions. In addition, the softest in-plane shear deformation in the QTA plot is along the $R_{cr}$ direction with $V(R_{cr})$ as low as around 2200 m/s, only half of those along the ridge and accordion directions. Similar to the LA case, the ratio of $V_{xy}/V(R_{cr})$, which is about 1.7, can be further tuned by strain with a significant increase with tensile $\varepsilon_x$ or compressive $\varepsilon_y$ strain as shown in Figure 3c. Therefore, the unique feature in the QTA is due to its puckered honeycomb lattice, which can be reversibly modulated by applying moderate mechanical deformation. Owing to the puckering structure, each neighboring two hexagons are connected by the orthogonal hinges[6] The loading of the $\varepsilon_x$ ($\varepsilon_y$) strains causes the minimum tension or shearing strains along the $R_{cr}$ and $R_{ha}$, ($R_{pr}$, and $R_{hz}$) directions, which are approximately diagonal between the armchair and zigzag directions. Therefore, the long-wavelength phonons, depending on the polarization of the atomic vibrations, are frozen under strain along these directions.

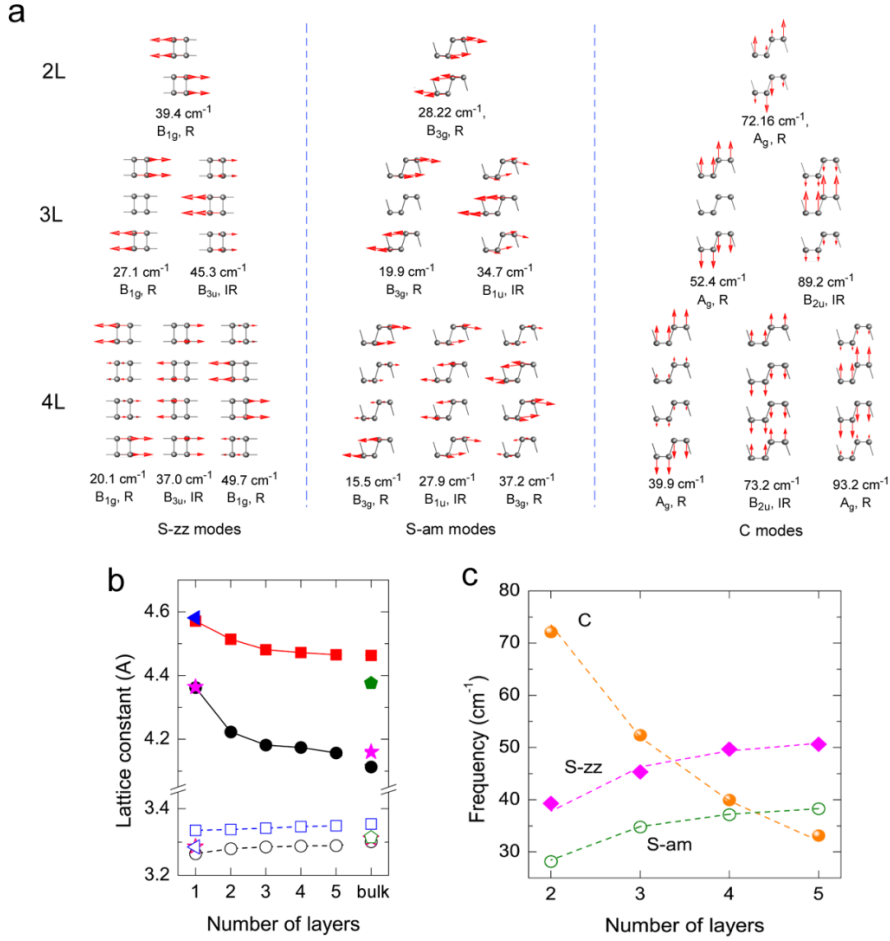

**Figure 4.** Normal mode displacements and frequencies of rigid shear and compression modes for few-layer phosphorene. **a,** Vibrating eigenvectors, frequencies, and symmetrical assignment of the shear modes parallel to the zigzag and armchair directions, denoted as S-zz and S-am, respectively, and compression (C) modes from 2L to 4L phosphorene. **b,** Evolution of lattice constants with thickness. The open (filled) symbols represent the lattice constant along the zigzag (armchair) direction. The circles are the results based on LDA calculation using Quantum-Espresso package. The squares, triangles and stars are the results from optB88, LDA and HSE06 hybrid functional calculations using VASP, respectively. The pentagons correspond to experimental measurement. **c,** Frequencies of the hardest S-zz and S-am modes, and the softest C mode as a function of the number of layers. The dashed lines are fitted according to the linear chain model.

We next explore low-frequency rigid vibrational modes for few-layer phosphorene. The stacking sequence of bulk BP in nature is AC stacking, where the out-of-plane phosphorus dimer in one layer is located directly above the center of the puckered hexagon of the neighboring layer. (see the inset of Figure 1a) In contrast to the naturally AB-staggered graphite and AA'-eclipsed favored h-BN, where at least one atom in each honeycomb layer is located below an atom in the neighboring layer, no phosphorus atom in each layer is found directly below an atom in the adjacent layer. This implies a different stabilization mechanism and rigid interlayer dynamics between few-layer phosphorene and perfect-honeycomb layers like graphene and BN.

Different from the intralayer vibrations in phosphorene, rigid-layer vibrations include the relative in-plane shearing and out-of-plane compressing displacement between neighboring layers (**Figure 4a**). Similar vibrations are found to be highly anharmonic and important for electron and phonon scattering in few-layer graphene[32] and $MoS_2$.[34] To examine the anisotropic behavior of these low-energy modes, we first examine the variation of the lattice constants with thickness (Figure 4b), which are fully relaxed by using both the LDA and optB88 functionals. Due to the well-known deficiency of the LDA and GGA functionals, the lattice constant tends to be underestimated by the LDA and overestimated by the optB88 functionals in comparison with the experimental value of bulk BP (see the pentagons in Figure 4b). Calculations using both LDA and optB88 methods show that the lattice constant along zigzag direction changes slightly, while that along the armchair direction clearly reduces with increasing thickness from 1L (4.36 Å, LDA result) to 5L (4.16 Å, LDA result) with a significant initial drop from 1L to 2L. Such a strong orientation-dependence of the lattice constant with the thickness, which is absent in graphene and MoS₂, indicates the anisotropic effect of the interlayer coupling.

**Table 1.** The frequency (ω) in unit of cm⁻¹ and Grüneisen parameter (γ) of the Raman and IR-active modes of phosphorene.

| | $B_{1u}$ | $B_{1g}$ | $B_{3g}^1$ | $A_g^1$ | $B_{3g}^2$ | $A_u$ | $B_{2g}$ | $A_g^2$ | $B_{2u}$ |
|---|---|---|---|---|---|---|---|---|---|
| ω (LDA) | 100.7 | 190.6 | 224.5 | 366.1 | 425.9 | 431.1 | 433.7 | 455.6 | 469.7 |
| ω (optB88) | 132.7 | 187.3 | 217.6 | 336.7 | 413.8 | 399.5 | 406.7 | 436.3 | 450.5 |
| ω (Ref. 21) | | | | 360.6 | | | | 438.9 | 469.0 |
| γ (LDA) | -8.63 | -0.01 | 0.06 | 0.58 | -0.54 | 1.07 | 1.01 | 0.32 | 0.92 |

The vibrational eigenvectors and the frequencies of the rigid shearing and layer-breathing modes for 1L to 4L are calculated by LDA method and illustrated in Figure 4a (See Figure S2 in the Supporting Information for the 5L case). For a given N, there are N-1 layer-breathing modes. The interlayer interaction also creates 2*(N-1) in-plane shearing modes, which are doubly degenerate in isotropic 2D materials like graphene and MoS₂. However, for phosphorene, as plotted in Figure 4a, this degeneracy is lifted, and the shearing between neighboring layers along zigzag or armchair direction with equivalent displacement patterns is accompanied with distinctly different frequencies with higher frequencies being along the zigzag direction. The N-1 rigid-layer modes, which are polarized along each direction, can be either Raman- or IR-active in an alternating sequence with the softest mode always Raman-active. The hardest mode corresponds to the rigid shift of innermost layers vibrating out of phase while the outer layers remain nearly fixed. The sensitivity of the frequencies of these modes on the thickness suggests an additional method capable of determining the number of layers by Raman or IR spectrum.

The relationship between the frequencies of the rigid-layer modes and the number of layers can be accurately analyzed by establishing a chain model, where only the nearest neighbor interaction is considered and each layer is treated as a single pseudoatom with a freedom of three. Similar models have been used for analyzing the shearing mode of graphene[32] and MoS₂,[34] where an angle-independent interlayer shearing motion is assumed. Here we show that the model can effectively describe the interlayer dynamics of phosphorene with an anisotropic character, where the interlayer coupling along different directions can be derived. The frequencies of the hardest shearing mode both along the armchair and zigzag directions, and the softest compression mode for N layers can be given by:

$$\omega_{\lambda z}(N) = \frac{1}{\sqrt{2}\pi c}\sqrt{\frac{\alpha_{\lambda z}}{\mu}}\sqrt{1 \pm \cos(\frac{\pi}{N})} \qquad (2)$$

where $\mu = 1.46 \times 10^{-6}$ kg/m² is the mass per unit area for the 1L phosphorene, $\alpha$ is the effective strength of the interlayer coupling, c is speed of light in unit of cm/s, and $\lambda$ is the type of mode which can be either a compression mode (denoted as "z") or shear mode along the zigzag (armchair) direction represented by "x" ("y"). The plus and minus signs correspond to the shear modes and compression modes, respectively. Figure 4c shows that the above model can fit our calculations perfectly with $\alpha_{xz}$, $\alpha_{yz}$, and $\alpha_{zz}$ being $3.7 \times 10^{19}, 2.1 \times 10^{19}$, and $1.4 \times 10^{20}$ N/m³, respectively. Based on the obtained α and considering the equilibrium distance (t=0.51 nm) between two layers in bulk BP, the modulus $C_{\lambda z} = \alpha_{\lambda z}t$ can be derived, with $C_{xz}$, $C_{yz}$, and $C_{zz}$ being 18.7, 10.6, and 70.8 GPa, respectively, which are in good agreement with experimental values of the shear moduli $C_{44}$ (21.3 GPa), $C_{55}$ (11.5 GPa), and stretching modulus of $C_{33}$ (70.0 GPa) through ultrasonic[56,57] and neutron scattering[58] measurements. Notably, the effective shear strength α and the shear modulus show a strong anisotropy, with the value along the zigzag direction being about 1.8 times larger than that along the armchair direction. Here $\alpha_{xz}$ and $\alpha_{yz}$ are 2.9 and 1.6 times larger than that of graphene ($1.28 \times 10^{19}$ N/m³),[32] respectively, and few-layer phosphorene tends to be harder to shear in plane than graphene. The shear strength of MoS₂ is found to be $2.7 \times 10^{19}$ N/m³ (Ref. 34), which is between the values along zigzag and armchair directions of phosphorene, whereas the compressive strength of the former ($4.2 \times 10^{20}$ N/m³ (Ref. 34)) is three times larger than that of phosphorene.

Previous studies on graphene and MoS₂ have shown that the rigid-layer modes are strongly anharmonic, accompanying with a strong umklapp scattering and electron-phonon interaction.[32,34] These low-energy modes have a significant effect on the electron and thermal transport due to the fully activated character at room temperature. While phosphorene possesses a puckered honeycomb structure, which is to a certain degree similar to graphene, the phononic properties are, however, dramatically different from the latter. In contrast to graphene, phosphorene has a strong orientation-dependent shear modulus, which indicates a strong anisotropy in the transport of phonons. For weak interlayer coupling, perturbation theory[59] gives an approximate estimation of the scattering rate (τ) for phonon modes propagating along $\lambda$ direction as $\tau_\lambda^{-1} \propto D_\lambda(\omega)K_\lambda^2/\omega^2$, where $D_\lambda(\omega)$ is the density of states, and $K_\lambda = \alpha_{\lambda z} * S$ is the effective interatomic force constant with S= 1.42×10⁻¹⁹ m⁻² being the area of the in-plane unit cell of mono-layer phosphorene. The effective interlayer coupling, given by $\alpha_{\lambda z}$ in few-layer phosphorene, is strongly anisotropic in the basal plane and much larger than that of graphene, suggesting a more pronounced interlayer interaction and thermal leakage normal to the layer direction and accordingly, a different strategy is needed for thermal management for phosphorene compared with graphene.

## 4. Conclusion

In summary, by examining the whole 2D Brillouin zone, we have theoretically demonstrated an extraordinary phononic anisotropy in phosphorene. We have found that this anisotropy is sensitively dependent on the magnitude and type of applied strain. From the low-frequency rigid-layer modes, we have found that there is a strong orientation-dependent coupling in few-layer phosphorene, with the coupling strength along the ridge direction being much stronger than that along the armchair direction. It is expected that these highly anisotropic and strain-sensitive features of phosphorene may offer new opportunities for fabricating novel flexible nanoelectronics, thermoelectronics, and phononics devices.

TOC figure

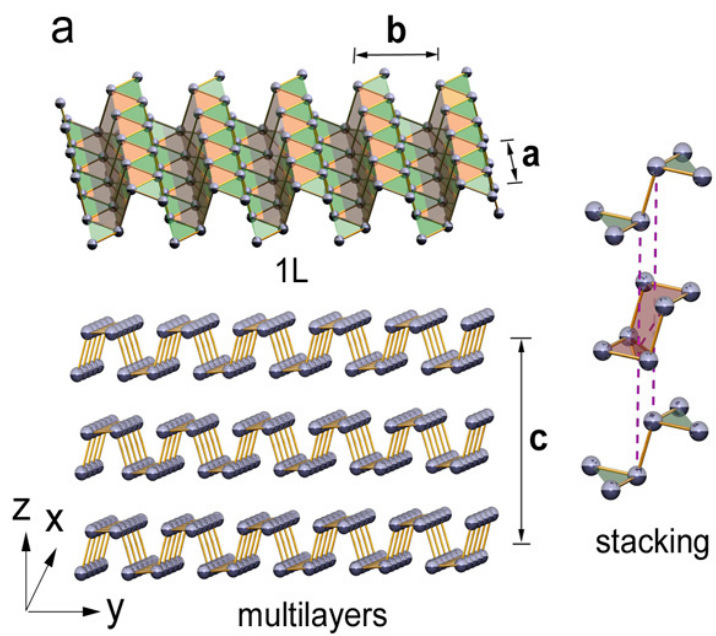